\documentclass[aps,prb,twocolumn,shortibliography,superscriptaddress, article]{revtex4-2}

\usepackage{epsfig}
\usepackage{epstopdf}
\usepackage{amsmath}
\usepackage{amsfonts}
\usepackage{amssymb}
\usepackage{hyperref}
\usepackage{bm}
\usepackage{makecell}
\usepackage{rotating}
\usepackage{hyperref}
\usepackage{multirow}
\usepackage{graphicx}
\usepackage{booktabs}
\usepackage{siunitx}
\usepackage{makecell}
\usepackage{textcomp}

\usepackage[normalem]{ulem}

\usepackage[dvipsnames]{xcolor}
\usepackage{soul}

\sethlcolor{Goldenrod}

\usepackage{graphicx}
\usepackage{dcolumn}
\usepackage{bm}
\usepackage{color}

\usepackage{tikz,xcolor,hyperref}

\definecolor{lime}{HTML}{A6CE39}
\DeclareRobustCommand{\orcidicon}{%
	\begin{tikzpicture}
	\draw[lime, fill=lime] (0,0)
	circle [radius=0.16]
	node[white] {{\fontfamily{qag}\selectfont \tiny ID}};
	\draw[white, fill=white] (-0.0625,0.095)
	circle [radius=0.007];
	\end{tikzpicture}
	\hspace{-2mm}
}

\foreach \x in {A, ..., Z}{%
	\expandafter\xdef\csname orcid\x\endcsname{\noexpand\href{https://orcid.org/\csname orcidauthor\x\endcsname}{\noexpand\orcidicon}}
}

\begin{document}

\title{Magnetic Frustration Enforced Electronic Reconstruction in Ni intercalated NbSe$_{2}$: Suppression of Electronic Orders}

\author{Ashutosh S. Wadge\orcidA}
\email{wadge@magtop.ifpan.edu.pl}
\affiliation{International Research Centre MagTop, Institute of Physics, Polish Academy of Sciences, Aleja Lotnik\'ow 32/46, PL-02668 Warsaw, Poland}

\author{Alexander Kazakov\orcidB}
\affiliation{International Research Centre MagTop, Institute of Physics, Polish Academy of Sciences, Aleja Lotnik\'ow 32/46, PL-02668 Warsaw, Poland}

\author{Xujia Gong\orcidC}
\affiliation{International Research Centre MagTop, Institute of Physics, Polish Academy of Sciences, Aleja Lotnik\'ow 32/46, PL-02668 Warsaw, Poland}

\author{Daniel Jastrzebski\orcidD}
\affiliation{International Research Centre MagTop, Institute of Physics, Polish Academy of Sciences, Aleja Lotnik\'ow 32/46, PL-02668 Warsaw, Poland}

\author{Bogdan J. Kowalski\orcidE}
\affiliation{Institute of Physics, Polish Academy of Sciences, Aleja Lotnik\'ow 32/46, PL-02668 Warsaw, Poland}

\author{Artem Lynnyk\orcidF}
\affiliation{Institute of Physics, Polish Academy of Sciences, Aleja Lotnik\'ow 32/46, PL-02668 Warsaw, Poland}

\author{Lukasz Plucinski\orcidG}
\affiliation{Peter Gr\"unberg Institut (PGI-6), Forschungszentrum J\"ulich GmbH, Wilhelm-Johnen-Stra\ss e, DE-52428 J\"ulich, Germany}
\affiliation{Institute for Experimental Physics II B, RWTH Aachen University, Sommerfeldstra\ss e 14, DE-52074 Aachen, Germany}

\author{Amar Fakhredine\orcidH}
\affiliation{Institute of Physics, Polish Academy of Sciences, Aleja Lotnik\'ow 32/46, PL-02668 Warsaw, Poland}

\author{Ryszard Diduszko\orcidI}
\affiliation{Institute of Physics, Polish Academy of Sciences, Aleja Lotnik\'ow 32/46, PL-02668 Warsaw, Poland}

\author{Marta Aleszkiewicz\orcidS}
\affiliation{Institute of Physics, Polish Academy of Sciences, Aleja Lotnik\'ow 32/46, PL-02668 Warsaw, Poland}

\author{Jedrzej Korczak\orcidJ}
\affiliation{International Research Centre MagTop, Institute of Physics, Polish Academy of Sciences,
Aleja Lotnik\'ow 32/46, PL-02668 Warsaw, Poland}

\author{Dawid Wutke\orcidK}
\affiliation{National Synchrotron Radiation Centre SOLARIS, Jagiellonian University, Czerwone Maki 98, PL-30392 Cracow, Poland}

\author{Marcin Rosmus\orcidL}
\affiliation{National Synchrotron Radiation Centre SOLARIS, Jagiellonian University, Czerwone Maki 98, PL-30392 Cracow, Poland}

\author{Rafal Kurleto\orcidM}
\affiliation{National Synchrotron Radiation Centre SOLARIS, Jagiellonian University, Czerwone Maki 98, PL-30392 Cracow, Poland}

\author{Natalia Olszowska\orcidN}
\affiliation{National Synchrotron Radiation Centre SOLARIS, Jagiellonian University, Czerwone Maki 98, PL-30392 Cracow, Poland}

\author{Carmine Autieri\orcidO}
\email{autieri@magtop.ifpan.edu.pl}
\affiliation{International Research Centre MagTop, Institute of Physics, Polish Academy of Sciences, Aleja Lotnik\'ow 32/46, PL-02668 Warsaw, Poland}
\affiliation{SPIN-CNR, UOS Salerno, IT-84084 Fisciano (SA), Italy}

\author{Andrzej Wisniewski\orcidP}
\affiliation{International Research Centre MagTop, Institute of Physics, Polish Academy of Sciences,
Aleja Lotnik\'ow 32/46, PL-02668 Warsaw, Poland}
\affiliation{Institute of Physics, Polish Academy of Sciences, Aleja Lotnik\'ow 32/46, PL-02668 Warsaw, Poland}

\date{\today}
\begin{abstract}
We investigate the single crystals of Ni$_{0.19}$NbSe$_2$, revealing that Ni intercalation profoundly alters the physical properties of NbSe$_2$. Magnetic measurements clearly show that the system is magnetically frustrated with antiferromagnetic ordering below 23.5\,K, with an irreversibility temperature near 10\,K, and a magnetic hysteresis with a small net magnetic moment. Overall, the system can be described as an inhomogeneous antiferromagnetic phase with magnetic disorder and magnetic frustration. 
We found two Curie-Weiss temperatures of -80\,K for the field in the {\it ab}-plane and -137\,K for the field out of plane, which are a consequence of anisotropic interactions in spin space and favor an orientation of the spin along the {\it c}-axis. Temperature-dependent resistivity shows a complete suppression of both charge density waves and superconducting order down to 300\,mK. Angle-resolved photoemission spectroscopy at 84\,K reveals a $\overline{\Gamma}$-centered electron pocket in Ni$_{0.19}$NbSe$_2$, which is absent in pristine NbSe$_2$. The electronic structure results show a shift of the van Hove singularity (VHS), which is the main cause of the suppression of the electronic orders.
These results align with recent theoretical predictions that Ni intercalation with cationic disorder favors frustrated antiferromagnetic stripe states, shifts the VHS and reconstructs the Fermi surface in NbSe$_2$. Our findings position Ni$_{0.19}$NbSe$_2$ within a magnetically frustrated, non-superconducting regime, highlighting how partial intercalation and disorder drive complex magnetic order and the Fermi surface reconstruction in low-dimensional quantum materials.
\end{abstract}


\maketitle
	
\section{Introduction}
NbSe$_{2}$, a prototypical layered transition metal dichalcogenide (TMD), has long served as a key platform for exploring the interplay between electronically ordered phases, particularly the coexistence and competition of charge density waves (CDWs) and superconductivity (SC) in low-dimensional systems \cite{simon2024transition}. Pristine 2H-NbSe$_{2}$ has a non-magnetic ground state and undergoes a CDW transition around 33\,K, followed by a superconducting transition around 7.3\,K \cite{bawden2016spin,lian2023interplay,neto2001charge,pasztor2021multiband}. In the ultra-thin limit, the ground state of NbSe$_2$ becomes even more sensitive to layer stacking sequences and interlayer Coulomb repulsion, which can modulate electronic correlations and suppress long-range order without the need for chemical doping \cite{xi2015strongly}. The interplay between these two states, though not fully antagonistic, remains sensitive to external perturbations such as strain \cite{kundu2024charge}, pressure \cite{moulding2020absence}, dimensionality \cite{xi2015strongly}, disorder \cite{dwedari2019disorder,chatterjee2015emergence} and intercalation \cite{naik2022evolution, toporova2020crystal,hauser1973effect}.
\begin{figure*}[htbp]
    \centering
    \includegraphics[width=1\linewidth]{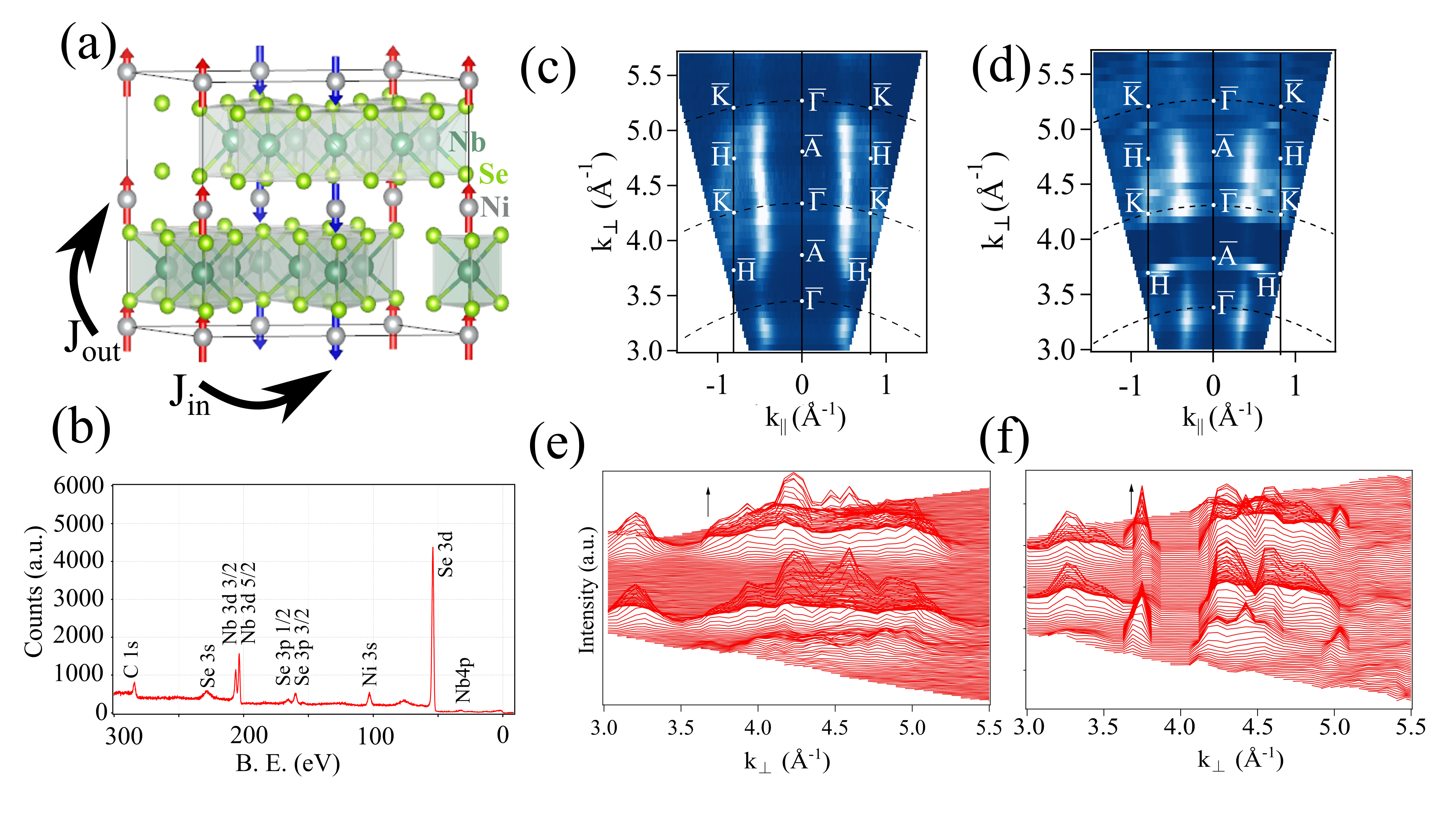}
	\caption{(a) Schematic of the proposed AFM2 magnetic structure in Ni$_{0.25}$NbSe$_2$, illustrating both out-of-plane ferromagnetic J$_{out}$ and in-plane antiferromagnetic J$_{in}$ coupling. The red and blue arrows represent the spin-up and spin-down of the Ni atoms. In the Ni$_{0.19}$NbSe$_2$ samples, there are Ni vacancies. (b) Core-level photoemission spectra of Ni$_{0.19}$NbSe$_2$, showing characteristic peaks from Ni, Nb, and Se orbitals. Out-of-plane (k$_{z}$) dispersion (at E$_{F}$) and corresponding waterfall plots for (c, e) pristine and (d, f) Ni-intercalated NbSe$_{2}$, revealing modifications to the electronic structure upon intercalation.}
   \label{fig:crystal_kz}  
\end{figure*}

Chemical intercalation offers a compelling route to perturb the ground state properties of TMDs, allowing insertion of atoms between van der Waals layers without significantly disrupting the in-plane lattice \cite{pervin2020second,erodici2023bridging,maksimovic2022strongly,popvcevic2022role,boucher2024intercalation}. In many cases, intercalants act primarily as charge donors or scattering centers, leading to modified carrier density and suppressed long-range order. However, when magnetic ions are introduced, they can profoundly affect the spin and electronic structure, potentially inducing novel correlated states \cite{friend1987electronic, edwards2024chemical}.
A literature survey has shown that the nature of the intercalants, whether non-magnetic or magnetic, plays a decisive role in determining the evolution of the electronic structure of NbSe$_{2}$ and related dichalcogenides. Alkali metal intercalation, such as Na, suppresses the CDW by eliminating soft phonon modes, while superconductivity remains at a reduced superconductive critical temperature \cite{lian2017first}. Magnetic ion intercalation causes more drastic effects: Fe intercalation induces a phase transition, disrupting CDW coherence through strong host–impurity interactions \cite{erodici2023bridging}. The Cr intercalation shows evidence of short-range magnetic clusters and Griffiths-like phases above the magnetic ordering temperature ($T_{order}$) \cite{ovchinnikov2022unbiased}.
Similarly, NiTa$_{4}$Se$_{8}$ exhibits strongly correlated itinerant magnetism competing with superconductivity \cite{maksimovic2022strongly}. Recently, Regmi et al. (2025) \cite{regmi2025altermagnetism} reported the discovery of altermagnetism in the layered transition metal dichalcogenide CoNb$_{4}$Se$_{8}$, which shows A-type antiferromagnetic ordering below 168\,K. Using a combination of experimental methods and density functional theory, they revealed momentum-dependent spin splitting despite zero net magnetization,
identifying the material as a “g-wave” altermagnet. Candelora et al. and Sakhya et al. \cite{candelora2025discovery, sakhya2025electronic} investigated ARPES spectra on altermagnetic CoNb$_{4}$Se$_{8}$ characterized by spin splitting without net magnetization. These phases arise from crystal symmetries rather than spin–orbit coupling, enabling non-relativistic spin splitting near the Fermi level. Altermagnets also exhibit promising functionalities, such as ultrafast optical switching \cite{de2025optical} and magnetic-field-tunable spin density waves \cite{candelora2025discovery}, highlighting their potential for novel spintronic applications.  Collectively, these studies demonstrate that magnetic intercalation profoundly modifies the electronic and magnetic ground states of NbSe$_{2}$-based materials, suppressing CDW and superconductivity while enabling emergent magnetic phenomena.

A recent study by Ramakrishna et al. \cite{ramakrishna2024synthesis} explored the structural and magnetic properties of Ni$_{x}$NbSe$_{2}$, reporting weak itinerant magnetism and a structural phase transition. However, key questions about magnetic frustration, the Fermi surface reconstruction and the suppression of collective states like CDW and superconductivity remain open. Notably, Gong \textit{et al.} \cite{gong2025tunability} theoretically explained the rich diversity of magnetic behaviors in Ni$_{x}$NbSe$_{2}$; it displays different magnetic ground states depending on the Ni content. For instance, Ni$_{0.33}$NbSe$_2$ exhibits altermagnetic order, which is in agreement with previous reports on altermagnetism for Ni$_{0.33}$NbS$_2$ below 84\,K \cite{Tenzin2025,PhysRevB.108.054418}. 
These examples underscore the sensitivity of the magnetic ground state to both carrier concentration, crystallographic environment and magnetic intercalant content.

In this study, we investigate the structural, magnetic transport, and electronic properties of Ni$_{0.19}$NbSe$_{2}$. 
We focus on the Ni concentration high enough to have long-range magnetic order, but far from the composition $\frac{1}{4}$ which gives the chance to have cation order with a 2$\times$2 reconstruction in the {\it ab}-plane \cite{de2025optical}. The Ni atoms sit only in the 2a Wyckoff position with an occupancy of $\frac{19}{25}$ shown by x-ray diffraction measurements and the stability of the Ni position in the 2a Wyckoff site was confirmed after structural relaxation performed using DFT.
We find that Ni intercalation completely suppresses both the CDW and SC phases of pristine NbSe$_{2}$ down to 300\,mK. Instead, magnetization measurements reveal frustrated antiferromagnetic behavior with an ordering temperature below 23.5\,K and an irreversibility temperature $(T_{irr})$: temperature at which ZFC and FC curves bifurcate, of around 10\,K, accompanied by a ferromagnetic hysteresis with a small net magnetic moment. 
Angle-resolved photoemission spectroscopy (ARPES), performed above the magnetic transition, reveals significant k$_{z}$-dependent modifications. The k$_{\parallel}$–k$_{\perp}$ plots shown in Fig. 1 (e–h) illustrate the redistribution of spectral weight at E$_{F}$  induced by Ni intercalation. First-principles calculations with Ni$_{0.25}$NbSe$_{2}$ reproduce the electron pocket at $\Gamma$, thereby confirming the Fermi surface reconstruction driven by intercalant-induced symmetry breaking. Together, these results demonstrate that Ni intercalation in NbSe$_{2}$ not only destroys the conventional electronic orders but also induces magnetic frustration and modifies the Fermi surface topology, revealing a new route to engineer correlated phases in low-dimensional TMDs.

The paper is divided as follows: the second Section is devoted to the experimental details and computational framework; the third Section presents experimental results on magnetic transport properties and ARPES band structure; and in the fourth Section, the authors draw their conclusions.

\section{Experimental Details and Computational Framework}
Single crystals of Ni$_{0.19}$NbSe$_{2}$ were synthesized via the chemical vapor transport method. High-purity elemental nickel, niobium, and selenium were mixed in stoichiometric proportions and placed into an alumina crucible. The crucible was then inserted into a quartz ampoule along with iodine as the transport agent at a concentration of 3 mg/cm$^{3}$. The ampoule was purged with high-purity argon for several minutes to eliminate moisture and oxygen, and subsequently sealed under high vacuum $\approx10^{-5}$ mbar.

The sealed ampoule was positioned in a three-zone horizontal tube furnace. The source zone was maintained at 860 $^{\circ}$C, while the growth (crystallization) zone was held at approximately 700 $^{\circ}$C. The temperature gradient was sustained for seven days to facilitate crystal growth. Upon completion, the furnace was allowed to cool naturally to room temperature. As a result, shiny, hexagonally shaped  Ni$_{0.19}$NbSe$_{2}$ single crystals were obtained and analyzed by x-ray diffraction as shown in the inset of Fig. S1 \cite{AWadgeSuppl}. Atomic force microscopy was carried out in tapping mode under ambient conditions to examine the surface morphology of both pristine and Ni-intercalated NbSe$_{2}$ samples. High-resolution atomic force microscopic images were acquired to resolve step features and surface texture at the nanoscale. Cross-sectional height profiles were extracted along representative line scans to quantify step heights and assess layer uniformity. In pristine NbSe$_{2}$, well-defined terraces and uniform step heights were observed, consistent with the layered structure. In contrast, Ni$_{0.19}$NbSe$_{2}$ showed increased surface roughness and irregular step edges, indicating morphological modifications due to Ni intercalation as shown in Figure S2 \cite{AWadgeSuppl}.

For magnetotransport measurements, single-crystalline specimens approximately 2 $\times$ 5\,mm in size were carefully cleaved and prepared. Standard four-probe electrical contacts were established using silver paint to ensure low-resistance (Ohmic), stable junctions. Longitudinal resistance was measured as a function of temperature during a controlled cooling process from room temperature down to $\approx$ 2\,K, utilizing a Quantum Design Physical Property Measurement System at a ramp rate of 2\,K/min. A low-frequency delta-mode excitation current in the range of 2.5 to 4\,$\mu$A was supplied via a Keithley 6221 current source, while corresponding voltage drops were detected using Keithley 2182A nanovoltmeters to ensure high signal fidelity. Additionally, superconducting transitions in Ni-intercalated samples were independently verified using a He-3 cryostat to access sub-2\,K temperatures with enhanced thermal stability.
Magnetic measurements were performed using a Quantum Design MPMS XL SQUID magnetometer equipped with the Reciprocating Sample Option, providing a magnetic moment sensitivity on the order of 10$^{-8}$\,emu. The single-crystal sample was mounted with the applied magnetic field oriented both parallel (in-plane) and perpendicular (out-of-plane) to the crystallographic basal plane.

Temperature-dependent magnetization was measured under an applied DC magnetic field of 100 Oe (0.01\,T) following standard zero-field-cooled warming (ZFC) and field-cooled cooling (FCC) protocols. Field-dependent magnetization loops were acquired after ZFC, at fixed temperatures of 5\,K, 15\,K, and 25\,K, with the field swept between $\pm$70\,kOe ($\pm$7\,T).

The ARPES experiment was conducted at the URANOS beamline of the SOLARIS National Synchrotron Radiation Centre in Cracow, Poland. Synchrotron radiation was provided by a quasi-periodic APPLE $\mathrm{II}$-type undulator, producing elliptically polarized photons in the energy range of 8–120\,eV. The flakes of NbSe$_{2}$ and Ni-intercalated NbSe$_{2}$ used for the measurements were exfoliated {\it in situ} using Kapton tape to obtain the clean surfaces. The exfoliation and measurements took place under ultrahigh vacuum conditions  $\approx4.9 \times10^{-11}$\,mbar. ARPES spectra were collected across a range of photon energies (20–120\,eV) and temperatures from 15-280\,K. Also, Fig. \ref{fig:crystal_kz}(b) shows core-level spectra; we have also collected LEED spectra as shown in Fig. \ref{fig:crystal_kz} (c, d).

We conducted calculations using the Vienna $ab$ $initio$ Simulation Package \cite{kresse1993ab,kresse1996efficiency} based on density functional theory (DFT) by the projector-augmented wave method \cite{kresse1999ultrasoft}. The exchange-correlation functional was described using the generalized gradient approximation (GGA) proposed by Perdew, Burke, and Ernzerhof \cite{perdew1996generalized}. The k-point grid was set to $10 \times 10 \times 5$ for the primitive cell and 
$12 \times 12 \times 3$, $5 \times 10 \times 5$ for different supercells, with a total energy convergence criterion of $10^{-4}$\,eV and a cutoff energy of 300 eV. The value of smearing $\sigma$ (a parameter which mimics finite-temperature broadening of the Fermi–Dirac distribution) significantly influences the magnetic property for U=0; here, we employ $\sigma$ = 0.1\,eV to simulate the non-magnetic phase. Ni$_{0.25}$NbSe$_{2}$ crystallizes in the space group  P6$_3$/$mmc$ (no. 194). The lattice parameters determined experimentally by XRD \cite{AWadgeSuppl} are $a=b=$3.4550\,{\AA} and $c$=12.3238\,{\AA}.
The 2 $\times$ 2 cell for the Ni$_{0.25}$NbSe$_{2}$ is obtained by double the in-plane lattice constant. Creating a supercell with 4 Ni atoms and removing one of them, we can obtain the stoichiometry Ni$_{0.19}$NbSe$_{2}$, which matches the experimental concentration of Ni. Therefore, to study the compound with Ni concentration 0.19, we need to further double the unit cell in the {\it ab}-plane and remove one atom, or double it along the {\it c}-axis and remove one atom. We study these two distinct types of vacancy defects; more details are reported in the supplementary materials. The lattice parameters are kept fixed while atomic positions are optimized. To capture the electron pocket near the Fermi level, we employ k-points grids of $18 \times 18 \times 4$ for the supercell doubled along the {\it c}-axis and $9 \times 18 \times 8$ for the supercell doubled in the {\it ab}-plane. 

\section {Results and Discussion}

\subsection{Magnetic Measurements Ni$_{0.19}$NbSe$_{2}$: Frustrated antiferromagnetic order with net magnetic moment}
\begin{figure}[h]
    \centering
    \includegraphics[width=\linewidth]{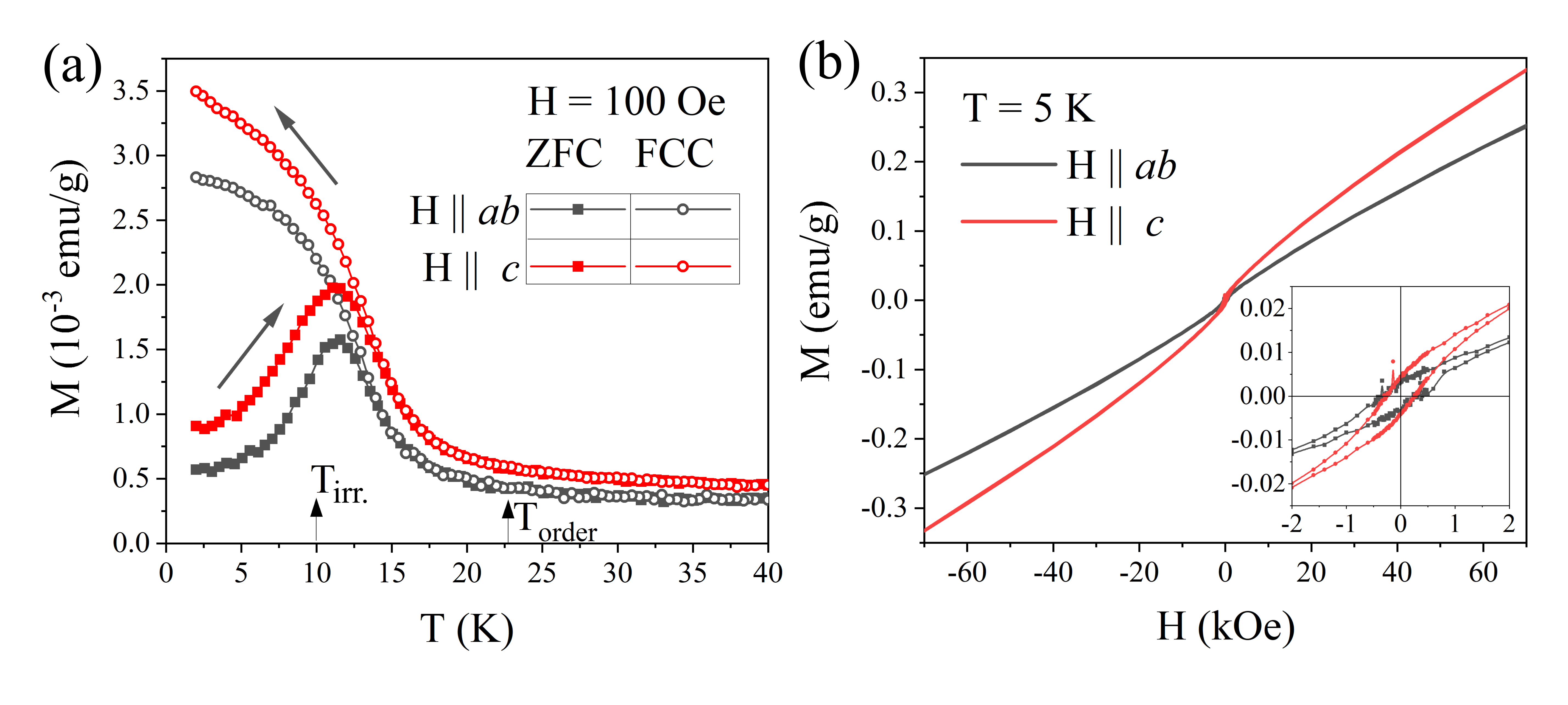}
	\caption{Magnetic measurements on Ni$_{0.19}$NbSe$_{2}$(a) Zero-field cooled warming (ZFC) and field cooled cooling (FCC) temperature dependence of magnetization, measured with magnetic field $H$ aligned either along the {\it c}-axis (red), or lying in the {\it ab}-plane (black). (b) Hysteresis loops obtained at 5~K for two field orientations. In the inset, we present a magnified view showing an unsaturated hysteresis loop.}
    \label{fig:Magnetization}
\end{figure}

Low-temperature $M-T$ dependencies measured along the {\it c}-axis and in the {\it ab}-plane (see Fig.~\ref{fig:Magnetization}a) exhibit typical behavior for the magnetically frustrated systems with competing ferromagnetic (FM) and antiferromagnetic (AFM) exchange interactions.

 \begin{figure}[htbp]
    \centering
    \includegraphics[width=\linewidth]{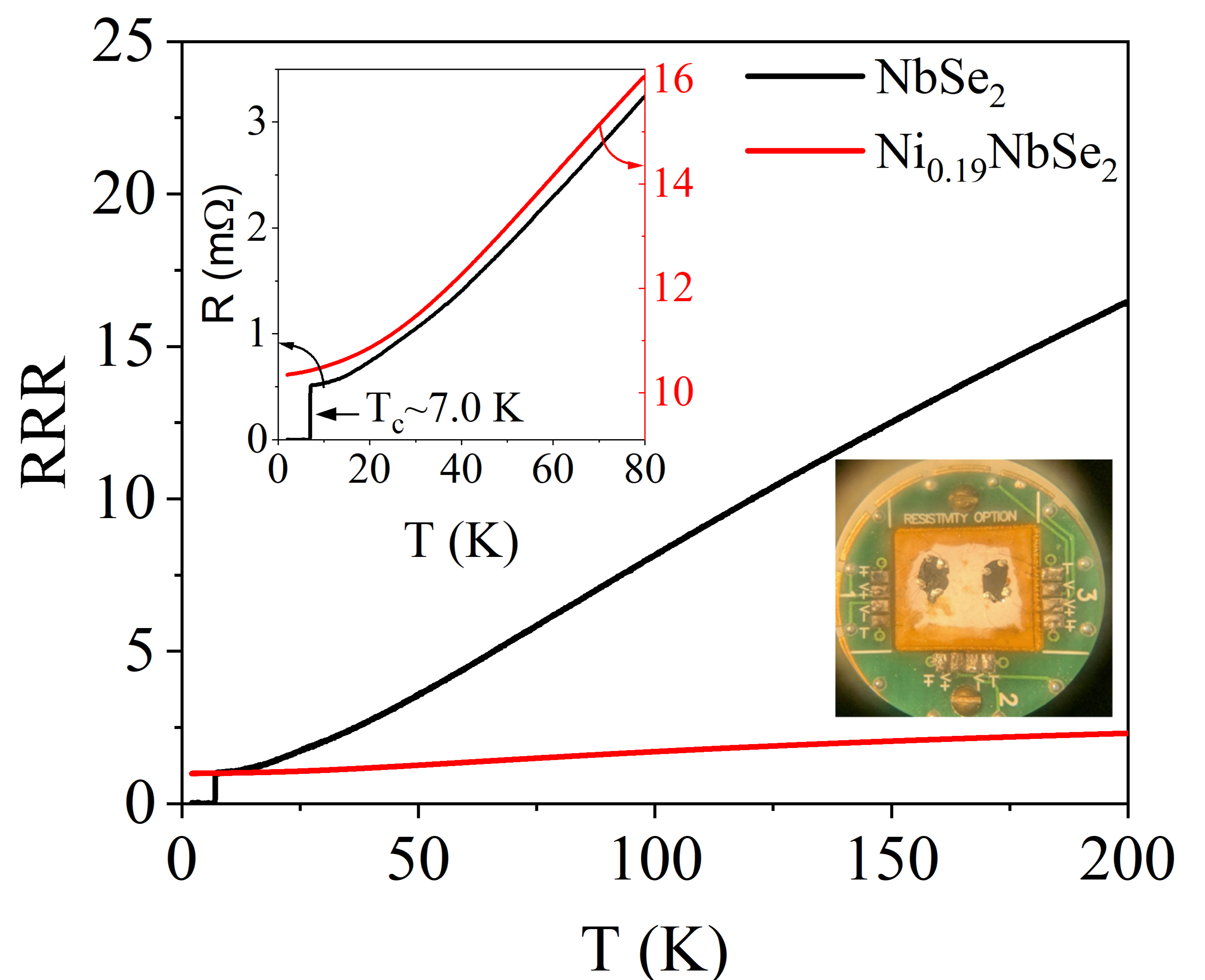}
    \caption{Temperature-dependent residual resistivity ratio (RRR) of pristine NbSe$_{2}$ and Ni$_{0.19}$NbSe$_{2}$. The pristine sample exhibits less pronounced CDW transition near 30 K and a clear superconducting transition around 7\,K, both of which are absent in Ni$_{0.19}$NbSe$_{2}$, indicating the suppression of collective electronic states upon Ni intercalation. The main inset shows a magnified view of the low-temperature region, and the secondary inset displays the actual crystal samples used in the transport experiments.}

    \label{Fig:transport} 
\end{figure} 

 \begin{figure*}[htbp]
    \centering
    \includegraphics[width=1\linewidth]{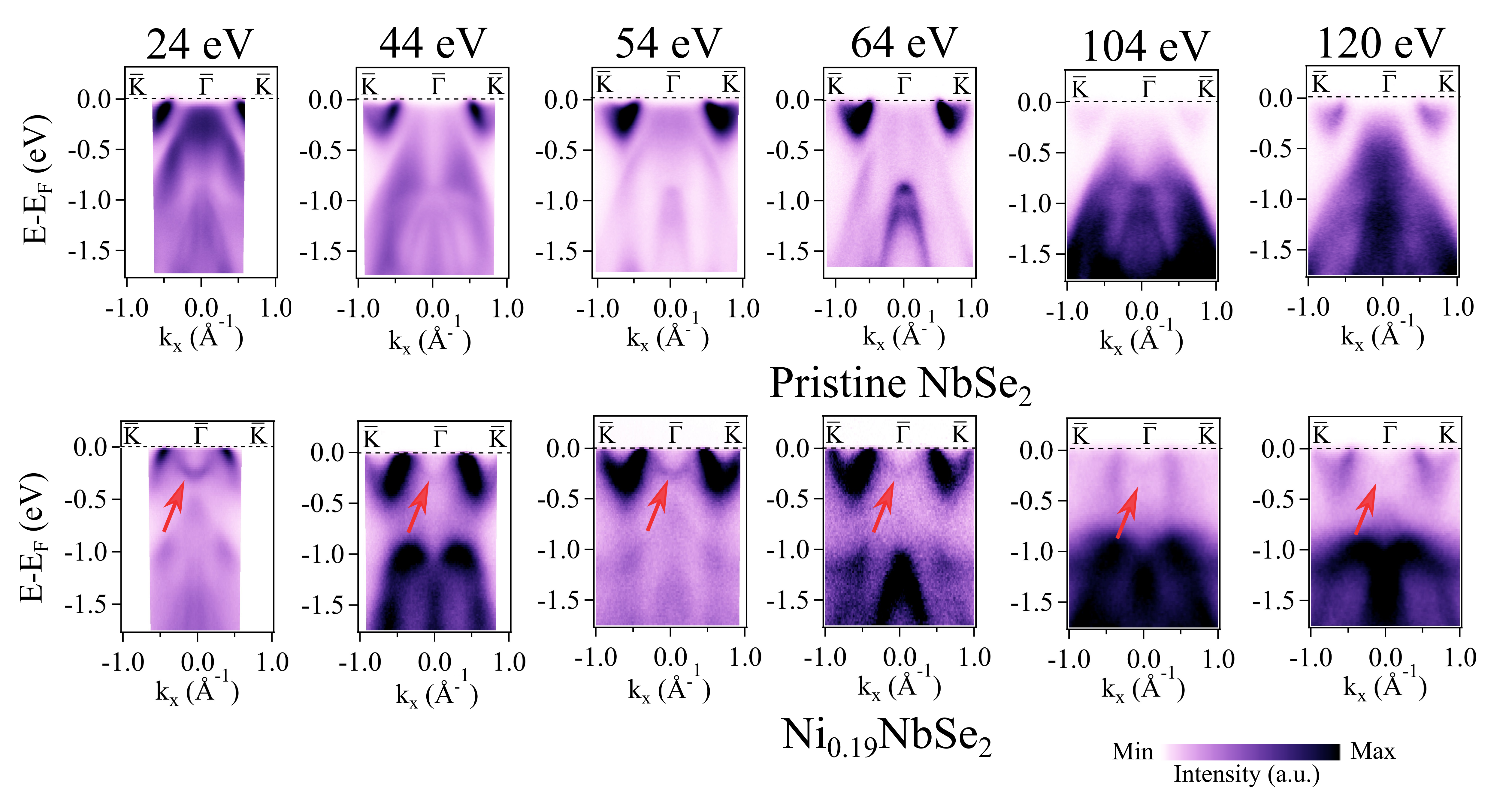}
    \caption{ARPES spectra at 85\,K along $\overline{K}$-$\overline{\Gamma}$-$\overline{K}$ path measured at photon energies 24, 44, 54, 64, 104 and 120 eV showing the comparison between (upper panel) pristine NbSe$_{2}$ and (lower panel) Ni$_{0.19}$NbSe$_{2}$ with extra electron pocket at the $\overline{\Gamma}$ indicated by red arrows. The colorbar reports the intensity in arbitrary units.}
    \label{Fig:energy} 
\end{figure*} 

At high temperatures, the material is in the paramagnetic phase and its magnetic susceptibility follows the Curie–Weiss law, ($\chi=\frac{C}{T-\theta_{CW}}$). To analyze the magnetic behavior in this regime, we fitted the high-temperature part of the inverse susceptibility with a linear function to extract the paramagnetic Curie–Weiss temperatures $\theta_{CW}$ \cite{Mugiraneza2022}. The fitted values reveal anisotropic Curie–Weiss temperatures of –80\,K in the {\it ab}-plane and –137\,K along the {\it c}-axis, confirming the presence of strong magnetic anisotropy. The temperature at which the experimental inverse susceptibility starts to deviate from linear Curie–Weiss behavior marks the onset of the crossover from the high-temperature paramagnetic regime to the magnetically correlated state, indicating the increasing importance of spin correlations \cite{chen2018short, lin2015unusual}. 
To model this anisotropic magnetism, we consider the XXZ Hamiltonian:
\begin{equation}
    H=\sum_{i<j}[J_{ij}^{\parallel}({S_i^x}{S_j^x}+{S_i^y}{S_j^y}) +J_{ij}^{\perp}{S_i^z}{S_j^z}]
\end{equation}
where S$_i^\alpha$ denotes the spin component at site $i$ and in the direction $\alpha$ Here, J$_{ij}^{\parallel}$ and J$_{ij}^{\perp}$ epresent the symmetric magnetic exchange interactions for the in-plane and out-of-plane components, respectively.  
The difference in the Curie–Weiss temperatures can be attributed to the difference between J$_{ij}^{\parallel}$ and J$_{ij}^{\perp}$, and/or to the second-order crystalline effect \cite{PhysRevB.7.3226,GREEDAN1984255}.
Negative values of the $\theta_{CW}$ indicate that AFM exchange interactions dominate in both cases. Higher absolute $\theta_{CW}$ values mean stronger magnetic interactions J$_{ij}^{\perp}$ compared to J$_{ij}^{\parallel}$ \cite{wang1971crystal, he2021anisotropic}, making the {\it c}-axis the easy-axis \cite{chevalier1985single}.
Anisotropic character of the $\theta_{CW}$ was also observed for other layered magnets \cite{chevalier1985single, armstrong2013anisotropic}.  

The presence of the frustration suppresses the magnetic transition temperature, and in the studied samples, the ordering temperature is $T_{order} = 23.5$\,K for both field orientations (see Figure S3 \cite{AWadgeSuppl}).
We can apply the Curie–Weiss law to determine its frustration index, $f$, which is defined as $f$=$\left| \frac{\theta_{CW}}{T_{order}}\right|$.
The values of f are 3.4 and 5.8 if we consider $\theta_{CW}$ for the {\it ab}-plane or the {\it c}-axis, respectively. Therefore, our system can be categorized as a moderately frustrated system\cite{Mugiraneza2022}. 

Below $T_{order}$, we observe bifurcation between zero-field cooled warming (ZFC) and field cooled cooling (FCC) measuring protocols, with an irreversibility temperature near 10 K. This behavior is typically indicative of the competing FM and AFM interactions, as expected in the present system \cite{gong2025tunability}. We also notice the susceptibility along the {\it c}-axis is slightly higher than in the {\it ab}-plane, suggesting that the easy axis lies along the {\it c}-axis, consistent with earlier calculations \cite{gong2025tunability}. Isothermal hysteresis loops (Fig.~\ref{fig:Magnetization}b), measured at low temperatures, further support the conclusions drawn from the temperature dependencies. Along the {\it c}-axis, magnetization aligns more easily along the {\it c}-axis than in the {\it ab}-plane. A coercive field of $\approx$300\,Oe was observed at 5\,K. Overall, magnetic properties indicate a complex magnetic ground state. However, detailed neutron scattering studies are necessary to determine the precise magnetic structure of the investigated crystals.


\subsection{Suppression of Charge Density Waves and Superconductivity}

Temperature-dependent residual resistivity ratio (RRR) measurements of Ni$_{0.19}$NbSe$_{2}$ and pristine NbSe$_{2}$ are presented in Fig.~\ref{Fig:transport}. The RRR is defined as the ratio of the resistance at elevated temperatures to the resistance at the lowest measured temperature above the superconducting transition $(T_{\mathrm{Low}} = 7.15\,K)$:
\begin{equation}
RRR = \frac{R(T_{\mathrm{High}})}{R(T_{\mathrm{Low}})}
\end{equation}
This ratio provides insight into the degree of impurity scattering and crystalline quality of the samples. The RRR of the Ni-intercalated NbSe$_2$ compound is lower than that of the pristine material due to the increased role of alloy and magnetic scattering. 
A pristine compound exhibits electronic instabilities at around 30\,K and 7.0\,K, corresponding to the less pronounced CDW and a significant SC transitions, respectively (Fig.~\ref{Fig:transport}). 
In contrast, the Ni-intercalated sample shows no sign of either transition, down to 300\,mK, though maintaining metallic behavior. 
This complete suppression of both CDW and SC orders is a hallmark of magnetic intercalation in TMDs. Previous studies of Co$_{x}$NbSe$_{2}$ \cite{chatterjee2015emergence} and Fe$_{x}$NbSe$_{2}$ \cite{naik2022evolution} have demonstrated that the introduction of localized magnetic moments introduces significant pair-breaking and momentum-space disorder, disrupting long-range electronic ordering. Our findings extend this understanding to Ni$_{0.19}$NbSe$_{2}$, where magnetic interactions evidently persist and strongly affect the electronic ground state.

\subsection{ARPES: Fermi surface reconstruction}

\begin{figure}[htbp]
    \centering
    \includegraphics[width=1\linewidth]{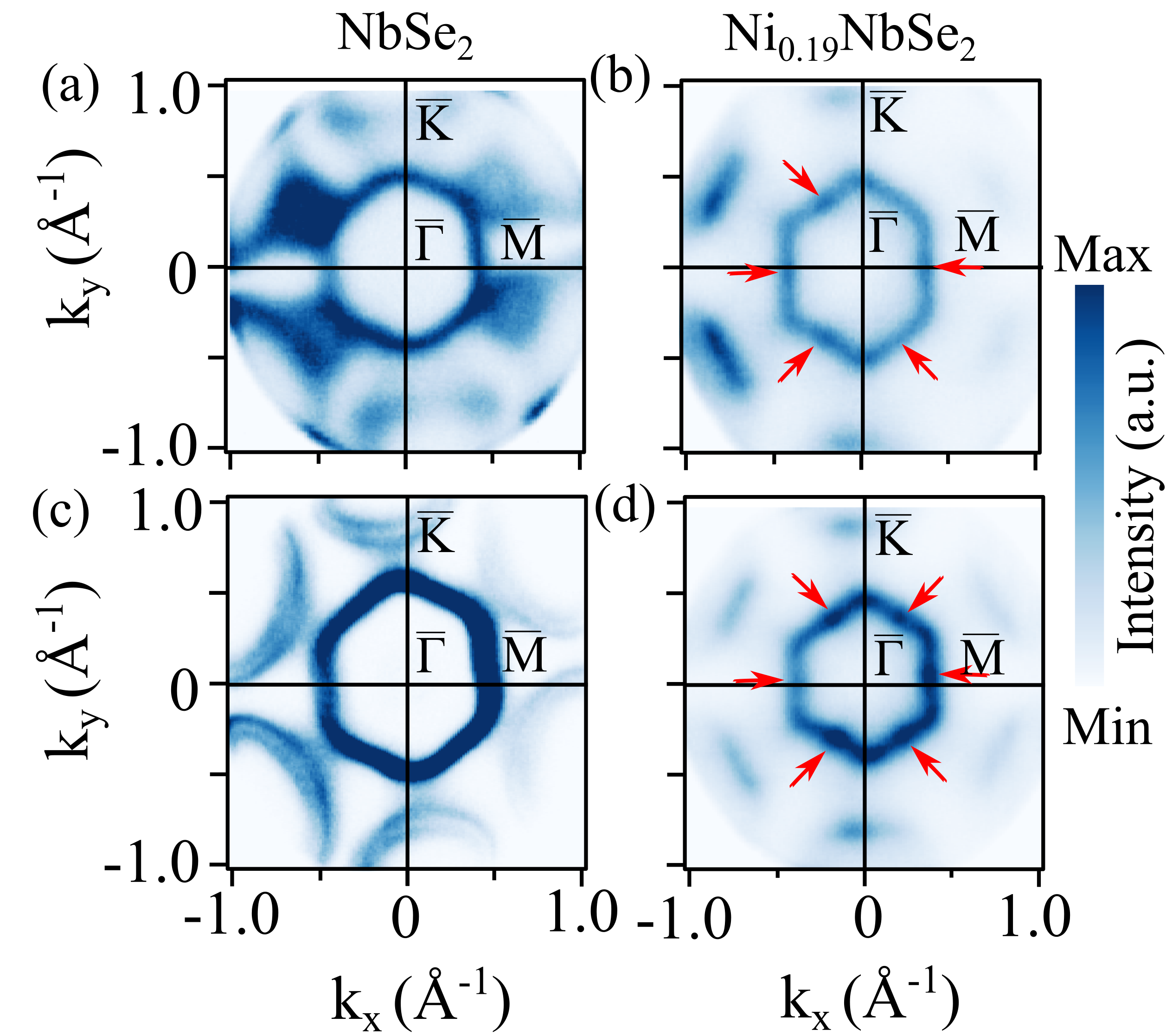}
    \caption{Fermi surface maps at 84 K of pristine NbSe$_2$ and Ni-interncalated NbSe$_2$ measured using ARPES. (a, b) Comparison of the Fermi surfaces for pristine NbSe$_2$ and Ni$_{0.19}$NbSe$_2$ resp. obtained from the sum of intensities of horizontal and vertical linear polarizations. The Ni-intercalated sample shows clear Fermi surface reconstruction, indicated by red arrows. (c, d) Fermi surface maps of NbSe$_2$ and Ni$_{0.19}$NbSe$_2$ obtained from the sum of intensities of left- and right-circular polarizations, further highlighting modifications in the electronic structure due to Ni intercalation (red arrows).}
    \label{Fig:FS} 
\end{figure}

\begin{figure}[htbp]
    \centering
    \includegraphics[width=1\linewidth]{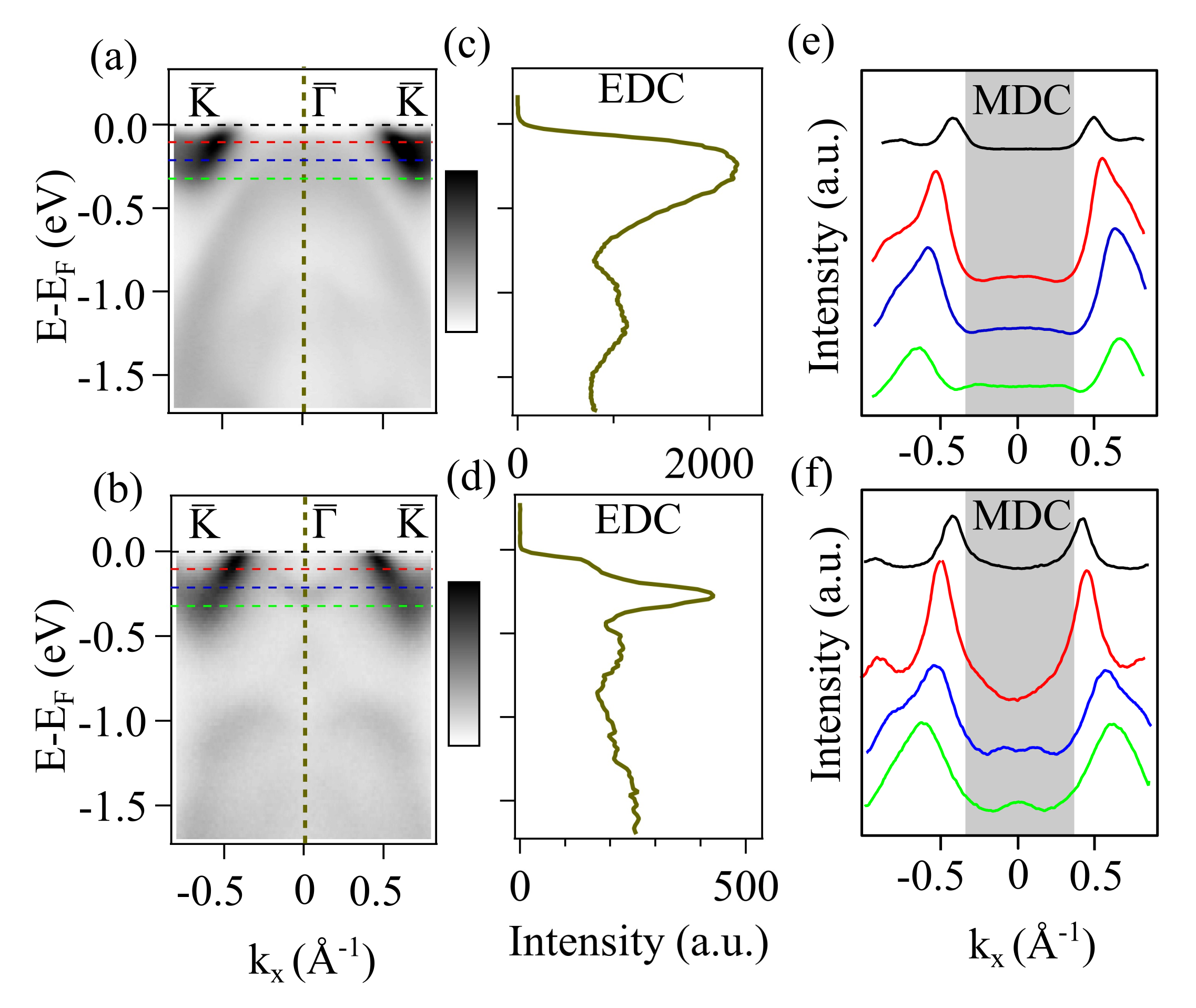}
    \caption{Comparison of electronic structure between pristine NbSe$_2$ and Ni$_{0.19}$NbSe$_2$ at 84\,K. 
(a, b) Band dispersion along the $\overline{K}-\overline{\Gamma}-\overline{K}$ direction for NbSe$_2$ and Ni$_{0.19}$NbSe$_2$, respectively. (c, d) Energy distribution curves (EDCs) at the $\overline{\Gamma}$ point reveal a significant near-$E_F$ intensity enhancement in Ni$_{0.19}$NbSe$_2$, consistent with the emergence of an additional electron pocket. (e, f) Momentum distribution curves (MDCs) at selected energies show distinct changes in intensity profiles, reflecting modifications to the electron band structure upon Ni intercalation. The selected energies using the same colors are reported in panels (a,b).}
    \label{Fig:EDC_MDC} 
\end{figure}

To probe the momentum-resolved electronic structure, we performed ARPES measurements on both pure and Ni-intercalated NbSe$_{2}$ at 84 K. Ni intercalation leads to a significant redistribution of spectral intensity, indicating substantial modification of the electronic structure (see Figures \ref{fig:crystal_kz} (e-h)). In addition, constant-energy maps presented in the Supplementary Material (Fig. S5) reveal the emergence of a $\overline{\Gamma}$-centered electron pocket with a band minimum at approx. 0.30 eV below E$_{F}$. The results, which exhibit energy dependence, are presented in Fig. \ref{Fig:energy}. The ARPES data for the pristine NbSe$_2$ are compared with the results in the literature \cite{kundu2024charge, bawden2016spin}. It also shows a hole band with a maximum at the $\overline{\Gamma}$ point, consistent with theoretical predictions \cite{calandra2009effect}. Using ARPES, we were able to observe portions of this hole band that lie below the Fermi level. The ARPES measurements on Ni$_{0.19}$NbSe$_{2}$ reveal a $\overline{\Gamma}$-centered electron-like pocket with a band minimum at approximately $-0.30$~eV, a feature that is completely absent in pristine NbSe$_{2}$. This pocket is not directly visible in the Fermi surface maps at $E_F$ (as shown in Fig. \ref{Fig:FS}) but becomes apparent in constant-energy contours at finite binding energies (Fig.~S6) and in the reconstructed 3D ARPES Fermi surface volume (Fig.~S7). Energy-momentum cuts along the $\overline{K}$-$\overline{\Gamma}$-$\overline{K}$ direction (see Fig. \ref{Fig:EDC_MDC}) also capture the dispersion of this pocket. 
These experimental results are consistent with DFT calculations 
(Figs.~S12 and S13~\cite{AWadgeSuppl}), which predict the emergence 
of such a $\overline{\Gamma}$-centered state upon Ni intercalation.
 
 
The linear- and circular-polarization composite maps of NbSe$_{2}$ and Ni$_{0.19}$NbSe$_{2}$ (Fig.\ref{Fig:FS} highlight the modifications of the electronic structure induced by Ni intercalation. The linear maps represent the sum of intensities from horizontal and vertical polarizations, while the circular maps represent the sum of left- and right-circular polarizations, all recorded under identical conditions. The emergence of an electron pocket at $\overline{\Gamma}$ cannot be explained by a simple rigid-band shift associated with electron doping but instead reflects the formation of a new hybridized state involving Ni and Nb $d$ orbitals, as shown in Fig. S8.
Further evidence of a broader electronic reconstruction is provided by the Fermi surface maps shown in Fig. \ref{Fig:FS}. Panels (a) and (b) compare the Fermi surfaces of pristine NbSe$_{2}$ and Ni$_{0.19}$NbSe$_{2}$, respectively, measured using combined linear horizontal and vertical polarizations to fully resolve the Fermi surface topology. In the Ni-intercalated sample (b), a clear reconstruction is observed, as indicated by the red arrows. The hexagonal-like contour around the $\overline{\Gamma}$ point is significantly reduced in size, and the six small electron pockets are shifted farther from the center, indicating substantial changes in the electronic structure. Panels (c) and (d), obtained with left and right circular polarizations, further highlight these reconstructed features. The consistent modifications observed across polarization modes confirm that Ni intercalation leads to robust alterations in the Fermi surface, possibly driven by enhanced electron correlations or symmetry-breaking effects such as band folding. A more descriptive comparison is shown in Fig. S4 \cite{AWadgeSuppl, zhang2011precise, rossnagel2001fermi, johannes2006fermi}.

\begin{figure}
    \centering
    \includegraphics[width=1\linewidth]{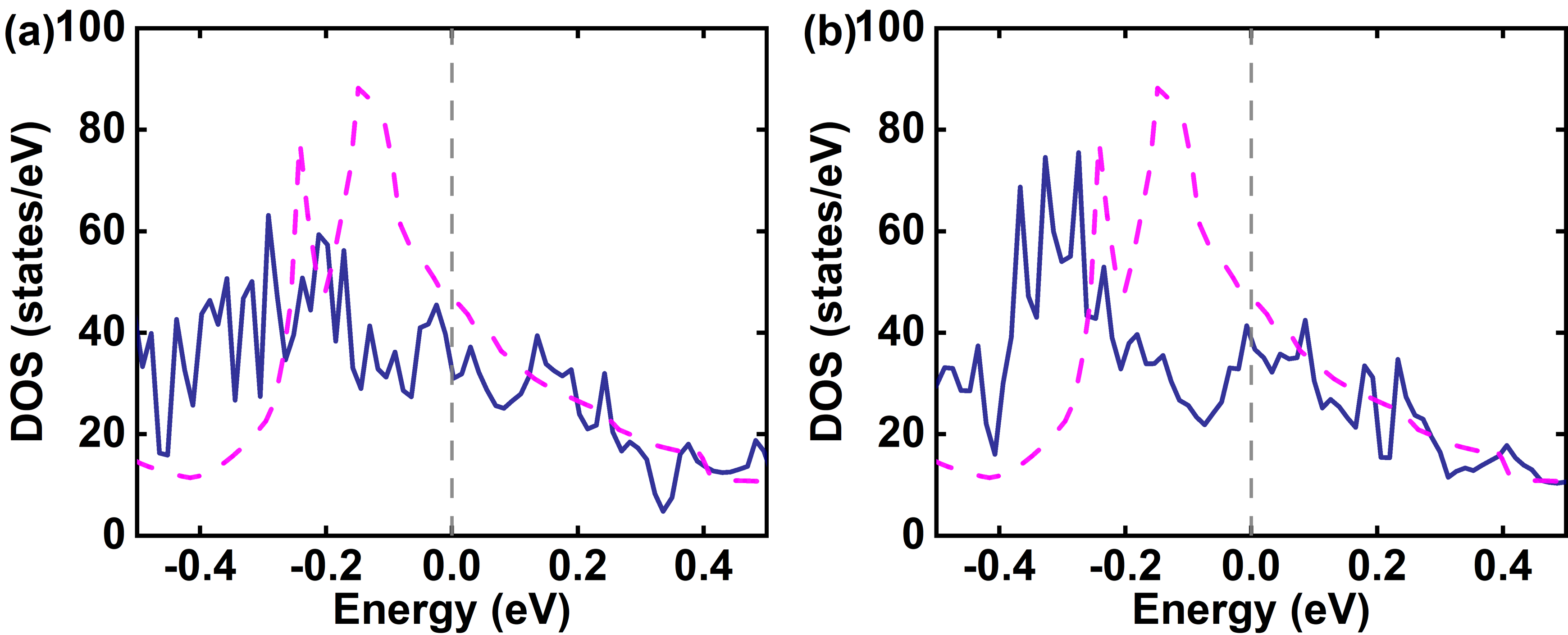}
    \caption{DOS of non-magnetic Ni$_{0.1875}$NbSe$_2$ calculated within GGA+SOC+U for (a) the layer without Ni and (b) the half-filled layer of Ni (blue solid line). To appreciate the shift of the VHS, we included the non-magnetic DOS for NbSe\textsubscript{2} (pink dotted line) in both panels. The NbSe$_2$ DOS is scaled to account for the different number of Nb atoms in NiNb$_4$Se$_8$.}
    \label{fig:DOS} 
\end{figure}

Band dispersion measurements at a photon energy of 48~eV, analyzed via momentum distribution curves (MDCs) and energy distribution curves (EDCs), are presented in Fig.~\ref{Fig:EDC_MDC}. The electron pocket $\overline{\Gamma}$ persists across multiple but not all photon energies (see Fig.~S5~\cite{AWadgeSuppl}), indicating its bulk origin. In addition to the formation of the electron pocket, the comparison of EDCs and MDCs between pristine and Ni$_{0.19}$NbSe$_2$ reveals significant spectral modifications as reported in Figs.~\ref{Fig:EDC_MDC}(c,d,e,f). In pristine NbSe$_2$, the peak of the DOS is 0.25 eV below the Fermi level, originating from the van Hove singularity at the M point. In the Ni-intercalated NbSe$_2$, the peak is 0.30 eV below the Fermi level; however, in this case, the peak is enhanced both by the van Hove singularity at the M point and by the additional contribution from the minimum of the $\Gamma$-centered electron pocket \cite{borisenko2009two}.  These changes are consistent with theoretical calculations of the DOS in the non-magnetic phase with two different configurations of Ni vacancies, shown in Fig.~\ref{fig:DOS}(a) and Fig.~\ref{fig:DOS}(b). In both cases, the VHS is displaced from the Fermi level relative to the pristine compound.
Ni intercalation introduces partial charge transfer to the NbSe$_2$ host, consistent with an effective Ni$^{2-\delta}$ valence ($\delta \approx x$) as reported in ~\cite{gong2025tunability, AWadgeSuppl}.

 \begin{figure}[htbp]
    \centering
    \includegraphics[width=1\linewidth]{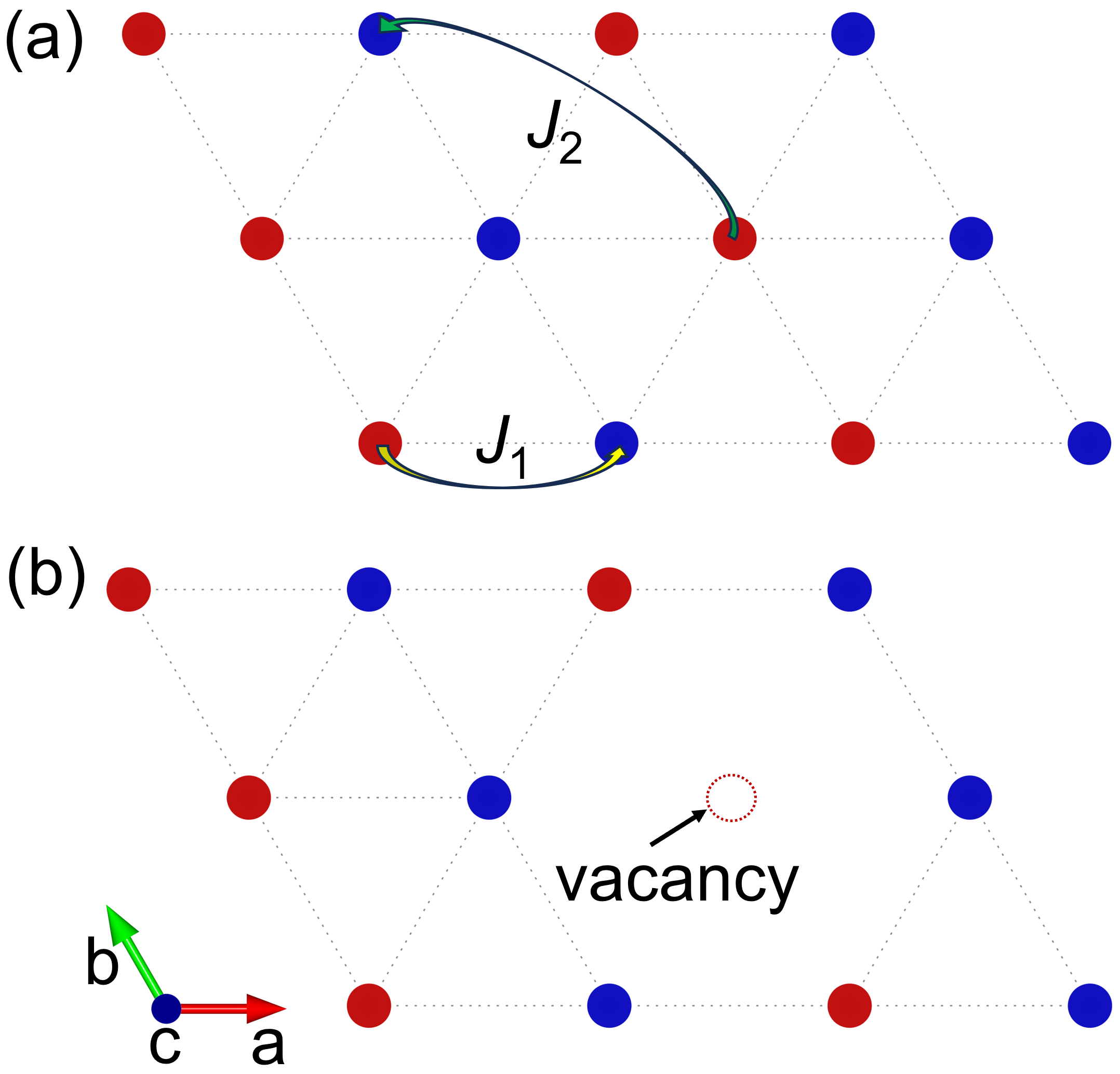}
    \caption{(a) Magnetic ground state configuration in {\it ab}-plane for the ordered phase of Ni$_{0.25}$NbSe$_{2}$, obtained from first-principles calculations \cite{gong2025tunability}. Red and blue balls represent Ni atoms with spin-up and spin-down, respectively. J$_{1}$ (named J$_{in}$ in Figure \ref{fig:crystal_kz} a) denotes the in-plane antiferromagnetic nearest-neighbor interaction, while J$_{2}$ represents the in-plane antiferromagnetic second-neighbor interaction. It is a stripe antiferromagnetic phase with time-reversal symmetry; the total magnetization is zero. (b) Magnetic ground state configuration in the {\it ab}-plane with a Ni vacancy representing the magnetic phase of Ni$_{0.19}$NbSe$_{2}$. Ni vacancies can induce magnetic disorder and result in a non-zero magnetization, which is responsible for the observed hysteresis.}

    \label{Fig:stripeAFM2} 
\end{figure}

A recent theoretical work by Gong \textit{et al.}~\cite{gong2025tunability} examines all four possible homogeneous magnetic phases in Ni$_{0.25}$NbSe$_2$: (1) a ferromagnetic phase (FM), (2) an altermagnetic phase (AM), (3) a stripe-like antiferromagnetic phase with antiferromagnetic J$_{out}$ (AFM1), and (4) a stripe-like antiferromagnetic phase with antiferromagnetic J$_{out}$ (AFM2), (see Fig.~\ref{fig:crystal_kz}a). 
The AFM2 phase is the energetically favorable one according to theoretical results\cite{gong2025tunability}. 
The altermagnetic phase presents relativistic weak ferromagnetism or weak ferrimagnetism depending on the N\'eel vector orientation; it is inherently a weak ferromagnet, as no orientation of the Néel vector results in a vanishing net magnetization. These weak ferromagnetism and weak ferrimagnetism come from the canting of the magnetic moments due to spin-orbit, which are always orthogonal to the N\'eel vector in this class of materials. 
Within density functional theory, the ground state of Ni$_{0.25}$NbSe$_2$ has been shown to be the AFM2 phase in the regime of weak electronic correlation. This is the physically relevant regime, owing to the metallic character of the compound, which screens the Coulomb repulsion \cite{gong2025tunability}. The introduction of Ni atoms leads to a significant Fermi surface reconstruction, including the formation of an electron pocket at the $\overline{\Gamma}$ point~\cite{gong2025tunability}. 
Our ARPES measurements directly observe such a pocket, which is compatible with the presence of AFM2 order. Additionally, supplementary DFT calculations (under the section: Theoretical explanation on disordered non-magnetic Ni$_{0.19}$NbSe$_{2}$) reveal that a non-magnetic state produces electron bands with an electron pocket too small compared with the experiments; therefore, the non-magnetic phase is discarded by our experimental data.

The homogeneous AFM2 phase of Ni$_{0.25}$NbSe$_2$ is reported in Fig. \ref{Fig:stripeAFM2}(a); the Ni atoms are on a triangular lattice with the first nearest neighbor J$_1$ antiferromagnetic and second nearest neighbor J$_2$ antiferromagnetic. While the antiferromagnetic J$_1$ is a source of frustration, the antiferromagnetic J$_2$ stabilizes the antiferromagnetic stripe phase if J$_2$/J$_1>$0.15.\cite{PhysRevB.92.140403} DFT results shown that AFM2 is the ground state; therefore, this should be attributed to a value of J$_2>$0.15J$_1$.
We have to stress that our system Ni$_{0.19}$NbSe$_2$ differs in the stoichiometry from Ni$_{0.25}$NbSe$_2$ and then falls below this homogeneity threshold and thus includes significant vacancy disorder on the Ni sublattice (See Fig. \ref{Fig:stripeAFM2}). The magnetic disorder originates from Ni vacancies, while the magnetic frustration arises from these vacancies and is further enhanced by the nearest-neighbor antiferromagnetic interaction $J_1$ on the triangular lattice. The presence of cationic disorder of the Ni atoms and Ni vacancy can push the system into the magnetically frustrated regime. 
Our magnetization data reveal a strongly negative Curie-Weiss temperature ($\theta_{CW} \approx -137$~K), a suppressed Néel temperature ($T_{order} \approx 23.5$~K), and a net ferromagnetic feature, collectively indicating a frustrated magnetic system with dominant antiferromagnetic interactions. The system presents no ferromagnetic saturated hysteresis loop, which discards the FM phase. Additionally, the small hysteresis loop discards homogeneous antiferromagnetic phases that preserve time-reversal symmetry. These observations allow us to exclude both the homogeneous ferromagnetic and homogeneous antiferromagnetic phases AFM1 and AFM2. Therefore, the only possible magnetic phase is an inhomogeneous antiferromagnetic (which can be named as ferrimagnetic) phase or an altermagnetic phase with weak ferromagnetism. Among the remaining two candidates, the disordered AFM2 phase offers a more consistent description of our experimental findings than the homogeneous altermagnetic phase. This disordered AFM2 phase also provides a natural mechanism for the observed suppression of both charge density wave CDW and SC orders. Stripe-like magnetic disorder disrupts Fermi surface nesting, a prerequisite for CDW formation, and introduces strong pair-breaking effects that destabilize superconductivity. Hence, our system lies close to the disordered AFM2 regime, where frustration, disorder, and electronic instability coalesce into a novel quantum state. 
The net ferromagnetic component observed in our system may originate from local inhomogeneities within the disordered AFM2 state. In particular, the magnetic disorder in the stripe antiferromagnetic phase can induce the formation of magnetic polarons, localized regions where the exchange field aligns neighboring spins, which in turn can stabilize a phase with a net magnetic moment experimentally observed~\cite{morera2024itinerant, van2022holes, bugajewski2025theory}. This mechanism offers a plausible route for the coexistence of frustration-dominated antiferromagnetic interactions and emergent ferromagnetism at low temperatures.

\section{Conclusions}

Our results reveal a moderately frustrated magnetic ground state of Ni$_{0.19}$NbSe$_{2}$ with dominant antiferromagnetic interactions, a magnetic hysteresis loop, and a large separation between the Curie-Weiss and the Néel temperatures (order temperature). These features eliminate homogeneous ferromagnetic and antiferromagnetic phases as a magnetic ground state.
The magnetic ordering temperature is 23.5\,K, however, we found two Curie-Weiss temperatures of -80\,K for the field in the {\it ab}-plane and -137\,K for the magnetic field out of plane, which derive from an anisotropic magnetic interaction in the spin space and favor an orientation of the spin along the {\it c}-axis. The frustration factor $f$ is 3.4 and 5.8 for both orientations, respectively, placing this compound in a regime of moderate frustration. Combined ARPES and theoretical analysis demonstrate a reconstructed Fermi surface with an electron pocket at the $\overline{\Gamma}$ point, consistent with the disordered antiferromagnetic state. The electronic structure results show a shift of the Van Hove singularity, which causes the suppression of the electronic orders, such as CDW and SC. The suppression of both CDW and SC orders further supports the presence of disorder-induced magnetism and electronic decoherence. Our findings firmly place Ni$_{0.19}$NbSe$_2$ within the inhomogeneous antiferromagnetic regime and underscore the broader potential of magnetic intercalation for engineering frustrated and incoherent electronic phases in layered quantum materials.\\

\begin{acknowledgments}
The authors thank F. Mazzola and M. Cuoco for useful discussions. This research was supported by the "MagTop" project (FENG.02.01-IP.05-0028/23) carried out within the "International Research
Agendas" programme of the Foundation for Polish Science, co-financed by the
European Union under the European Funds for Smart Economy 2021-2027 (FENG). Research at the National Synchrotron Radiation Centre SOLARIS is supported by the Ministry of Science and Higher Education, Poland, under contract no. 1/SOL/2021/2. C.A. acknowledges support from PNRR MUR project PE0000023-NQSTI.
We further acknowledge access to the computing facilities of the Interdisciplinary Center of Modeling at the University of Warsaw, Grants g91-1418, g91-1419, g96-1808 and g96-1809 for the availability of high-performance computing resources and support.  We acknowledge the access to the computing facilities of the Poznan Supercomputing and Networking Center, Grants No. pl0267-01, pl0365-01 and pl0471-01.
\end{acknowledgments}

\section*{Author's Contributions} A.S.W. grew the samples with assistance from D.J. and J.K.; A.K. performed the electron transport measurements and analyzed the magnetization data. A.L. conducted SQUID experiments. X.G., A.F., and C.A. carried out the DFT calculations. R.D. performed X-ray diffraction measurements and analysis. L.P. carried out calculations of orbital contributions and spin texture. M.A. conducted atomic force microscopy experiments. ARPES measurements were performed by A.S.W., B.J.K., D.J., and A.W., with support from M.R., D.W., R.K., and N.O.; A.S.W. analyzed the ARPES data and wrote the original draft with input from C.A. and A.K.; A.S.W. and A.W. conceived the project. A.W. supervised the entire project. All authors contributed to discussions and provided feedback on the manuscript.

\section*{Data Availability}
The data that support the findings of this study are available from the corresponding author upon reasonable request.
\medskip


\bibliography{references}  

\end{document}